 \documentclass[preprint2]{aastex}

\shorttitle{Star Formation \& Gravitational Microlensing}
\shortauthors{Gil-Merino \& Lewis}

\begin{document}


\title{Seeing Star Formation Regions with Gravitational Microlensing}

\author{Rodrigo Gil-Merino\altaffilmark{1} \& Geraint F. Lewis\altaffilmark{2}}
\affil{Institute of Astronomy, School of Physics, 
University of Sydney, NSW 2006, Australia}
\altaffiltext{1}{rodrigo@physics.usyd.edu.au}
\altaffiltext{2}{gfl@physics.usyd.edu.au}

\begin{abstract}
We qualitatively  study the  effects of gravitational  microlensing on
our view of unresolved  extragalactic star formation regions.  Using a
general gravitational microlensing  configuration, we perform a number
of simulations that reveal that  specific imprints of the star forming
region are imprinted, both photometrically and spectroscopically, upon
observations.   Such observations  have  the potential  to reveal  the
nature and size  of these star forming regions,  through the degree of
variability observed  in a monitoring campaign, and  hence resolve the
star formation regions  in distant galaxies which are  too small to be
probed via more standard techniques.
\end{abstract}


\section{Introduction}

Gravitational microlensing is now a well established technique for the
investigation  of the  distribution of  compact (dark)  matter  in the
universe.   Furthermore, it  also provides  a powerful  tool  to study
unresolved  sources, such as  in the  case of  the structure  of QSOs,
through  temporal   differential  magnification  (e.g.    Yonehara  et
al. 1998).

From an  observer's point of  view, gravitational microlensing  can be
naturally  divided in  two different  regimes.  In  the  case of Galactic
microlensing, the optical  depth is low and a  single star microlenses
another star within the Galactic halo or in one of the galaxies in the
Local  Group (Paczy\'nski  1986a).   With Extragalactic  microlensing,
where the light from a  distant quasar shines through a closer galaxy,
the optical  depth is roughly unity  and many stars  contribute to the
overall  microlensing effect (Paczy\'nski  1986b).   This paper
considers this latter regime, were the source region is populated by a
number of hot, young stars in a star forming region.  Such a situation
will occur  in strongly  lensed, multiply imaged  systems, such  as the
multiple images  seen in galaxy  clusters (Mellier 1999), or  the case
where a  isolated galaxy gravitationally lenses a  more distant galaxy
(i.e. Warren et al 1996).

In a similar vein, Lewis \&  Ibata (2001) investigated the effect of a
cosmological distribution of compact objects on the surface brightness
distributions   of  galaxies   at  $z$$<$0.5,   considering   a  small
microlensing  optical  depth  ($\leq$0.04)  and they  determined  that
low-level  fluctuations  in  surface  brightness of  $\sim$2\%  should
result.   Lewis  et al.   (2000)  extended  that  analysis to  distant
galaxies observed through galaxy  clusters, assuming dark matter to be
composed of compact objects.  Focusing upon Abell~370 as a case study,
concluding  that  for  low-luminosity ($\sim$10$^4$L$_\odot$)  stellar
populations would  show rapid fluctuations exceeding 10\%  of the mean
in the highest cases.

In  this contribution  we address  the question  of  what microlensing
signatures should be apparent in the case of part of a galaxy which is
lensed by another  galaxy.  In particular, if the  lensed parts of the
source  galaxy are  regions  of star  formation,  highly dominated  by
young, massive stars.  Such a  situation was recent presented by Smith
et  al.    (2005)  who  reported   the  discovery  of  a   new  strong
gravitationally lensed system, with an elliptical galaxy acting as the
lens.  The  lens galaxy in this  system is at  redshift $z=0.0345$ and
the source, proposed to be a star formation region, is at $z\sim0.45$,
with the arcs formed by the gravitational mirage showing `knots' of an
extreme  blue  color  of  $B-I_c=1.1$  (extinction  corrected).   This
discovery poses the  idea that microlensing in the  multiple images of
these systems  might be able to  distinguish the type  of source stars
involved in the mirage and help in the interpretation of its nature.


Within the  context of gravitational lensing, a  star formation region
would appear as  a non-uniform source, composed of  a number of bright
points in a more  extended background. Hence, the microlensing imprint
of such  a source  should show quite  a different  variability imprint
from  the  uniform   sources  typically  considered  in  gravitational
microlensing experiments. The nature of this imprint is the basis of 
this current contribution.

\section{Microlensing simulations}\label{sim}

For the purpose of this study we performed microlensing simulations by
means  of  ray-shooting techniques  (Paczy\'nski  1986b, Schneider  \&
Weiss 1987, Kayser  et al. 1986, Wambsganss 1990,  Witt 1993, Lewis et
al.   1993).  To  compute magnification  patterns, one  has  to select
certain values  for the  convergence ($\kappa$), which  represents the
gravitational  potential due  to matter  in  the beam,  and the  shear
($\gamma$), which  is the  perturbation to the  beam due to  the large
scale distribution  of matter.  Typically, these  parameters are drawn
from a  lens model for a  particular system. For  this study, however,
representative values of $\kappa=0.55$ and $\gamma=0.55$ are employed,
following Schechter et al.   (2004), although other combinations would
illustrate  the situation  equally; $kappa$ here includes also any form 
of compact dark matter, the effects of an smooth dark matter component 
are described in Schechter \& Wambsganss (2002).  Since  high resolution  maps are
required,  we used  a receiving  field of  2 Einstein  radii\footnote{
The Einstein radius is defined in the source plane as $ER=sqrt{(4GM/c^2)
(D_{s}D_{ls}/D_{l})}$, where $M$ is the mass of the microlens, $D$ is the 
angular distance to the source (s), the lens (l) and between the lens and 
the source (ls), c is the velocity of light and G the gravitational 
constant},  
covered  by  a  $2048^2$   pixels  area.   The
microlenses were  randomly distributed and  selected to have  the same
mass, $M_{\mu  lens}=1 M_{\odot}$.  Again,  the selection of  the mass
range is arbitrary for our  purposes. However, it is important to note
that rather  than covering  a large area  in the simulations,  the key
point remains in the resolution of the magnification patterns, because
we are  interested in small  flux changes from  pixel to pixel,  so we
also selected  a high  number of  rays that resulted  in over  700 per
pixel on average.

The  next step in  the simulations  is introducing  the effect  of the
source.  To  do this,  we assumed  a source plane  at $z=0.5$  and two
different  sizes  of  $0.1$   and  $0.5$~Einstein  radii  (ER),  which
corresponds to a physical  size of $0.02$~pc and $0.1$~pc respectively
at that  distance for the standard  $\Lambda$CDM cosmoslogy.  Although
star  formation  regions  might   be  larger  than  the  bigger  size
considered, these  two examples will illustrate  the different effects
due to  their sizes  and could  be seen as  clumps of  star formations
within larger regions (compact and ultra-compact H~{\small II} regions
as indicators of star formation might be $<$0.1~pc, see e.g. Giveon et
al. 2005 and references therein).  We also assumed that our lens plane
is at $z=0.04$  (following the case of Smith  et al. 2005).  Depending
on the  stellar density of the  source region, the number  of stars in
that  region can vary  from just  a few  up to  hundreds.  Considering
first  the $0.1$~Einstein  radii  region, we  `built' three  different
sources: one containing  8 stars, another containing 80,  and the last
one as  a uniform source, i.e.,  containing one star per  pixel in the
region (the number  of stars are not representative  of any particular
region, and have been chosen arbitrarily).


\begin{figure}[h]
\epsscale{1.00}
\plotone{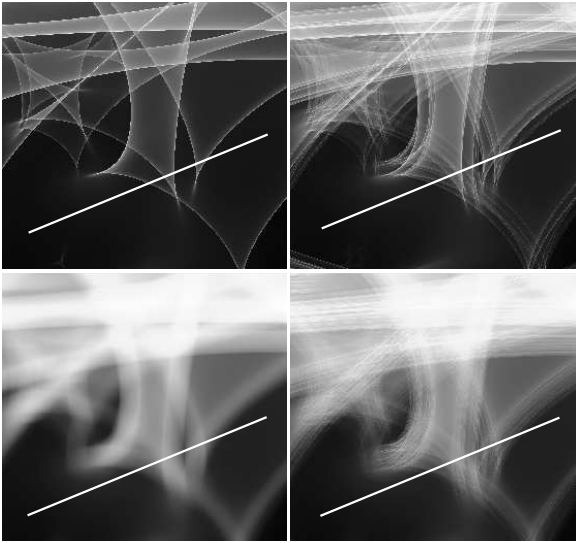}
\caption{The  original  magnification pattern  (upper  left panel)  is
convolved with a field of 8 stars (upper right panel), 80 stars (lower
right  panel)   and  a  `uniform  source'  (lower   left  panel).  The
magnification map is 1 ER on a side.}
\label{fig1}
\end{figure}


The   results   for  the   first   region   size   are  displayed   in
Fig.~\ref{fig1}.   The   upper   left-hand   pannel   corresponds to a
$\sim$1~ER$^2$ region of the  original magnifcation pattern. The upper
right-hand pannel is the magnification pattern convolved with a region
of $0.1$~ER containing 8 stars.  The lower right-pannel is the same as
the previous one but containing  80 stars.  The lower left-hand pannel
is the magnification pannel convolved  with an `uniform' source of the
same physical size.


\begin{figure}[h]
\epsscale{1.00} \plotone{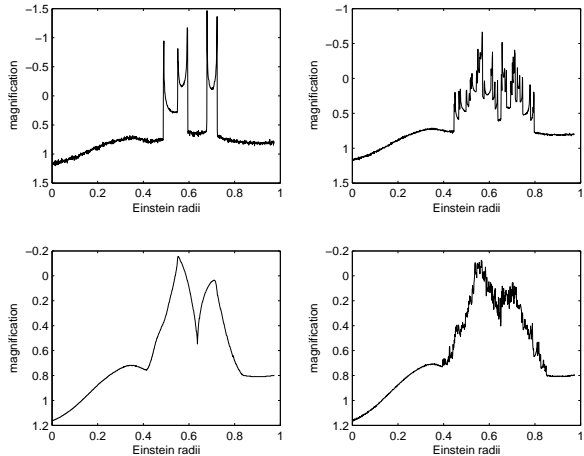}
\caption{The  corresponding  synthetic  light  curves  to  the  tracks
plotted  in Fig.~\ref{fig1}.  When  a few  stars  dominate the  source
region (upper-right pannel), the variability can be easily distinguish
from a `uniform' source (lower-left pannel).}
\label{fig3}
\end{figure}

\begin{figure}[h]
\epsscale{1.00}
\plotone{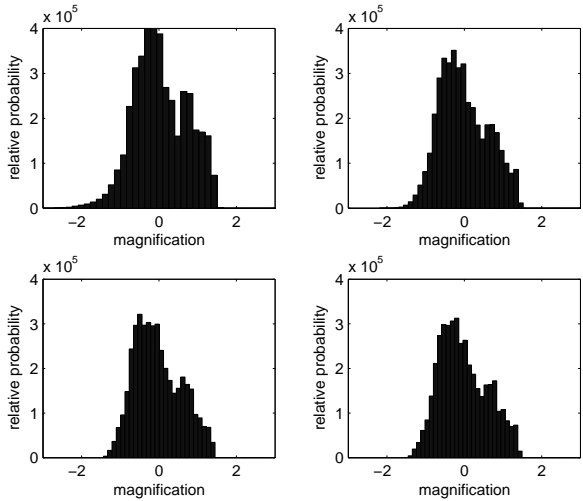}
\caption{The relative  probability distributions of  the corresponding
pannels in Fig.~\ref{fig1}.}
\label{fig2}
\end{figure}

In all the panels the same  track has been drawn, in order to compared
the  synthetical light  curves to  each other;  these are  depicted in
Fig.~\ref{fig3}.  The  light curves  are $\sim$1~ER long,  showing the
different  expected   fluctuations  corresponding  to   the  different
scenarios.   The  magnification  distributions for  the  magnification
patterns corresponding to the  different panels in Fig.~\ref{fig1} are
shown in  Fig.~\ref{fig2}. Clearly the  number of stars in  the region
has a significant  influence on the resulting light  curve; in effect,
the  presence  of  each  star  produces  a  ``shift-and-add''  to  the
mangification map,  greatly increasing the number  and overall density
of  caustics.  This  is reflected  as  additional peaks  in the  light
curve.  As the number of stars in increased to 80, some of the caustic
structure  has begun  to wash  out, leaving  small  scale fluctuations
superimposed on  a more gentle  background, whereas the  smooth source
(which can be  thought of as a very high density  of stars) has washed
out all small scale detail.


\begin{figure}[ht]
\epsscale{1.00}
\plotone{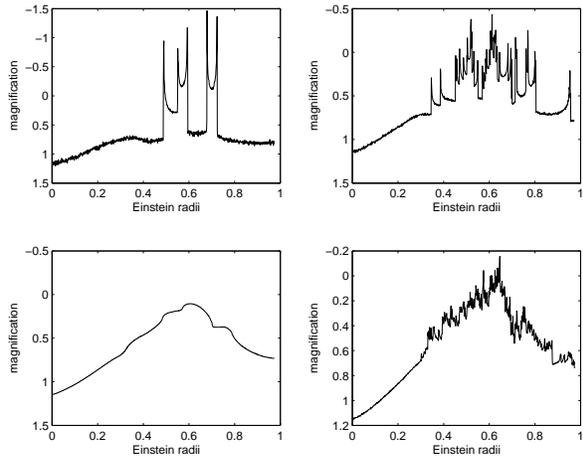}
\caption{Light curves considering a
region with a  size of $0.5$~ER.}
\label{fig3b}
\end{figure}

In Fig.~\ref{fig3b}, for comparison,  we  consider a
region  size of  $0.5$~ER  containing also  8  stars, 80  stars and  a
`uniform' source in the same manner as in Fig.~\ref{fig1}. 
The corresponding  track shows  a completely  different light
curves  compare to  Fig.~\ref{fig3} although  their positions  are the
same,  due to  the new  caustic  structure of  the magnification  maps
according to the different size of the region considered.


The  interpretation  of  these   figures:  if  the
magnification  pattern is  convolved with  a `uniform'  source profile
(Fig.~\ref{fig1}, lower  left panel) the  result is always  a smoother
pattern,  with smooth transitions  in the  value of  the magnification
from  pixel to  pixel;  if the  source area  is  made of  a number  of
point-like objects,  the convolution will show  many caustics slightly
shifted  one another,  with no  smooth transition  between  them. This
translates  into  a  rapid  variability  in  the  lightcurves  of  the
corresponding source.  Also, the size of the  regions considered plays
an  active  role   in  the  final  imprint  of   microlensing  in  the
observational light curves.

\section{Applications and Discussion}

The application  of the simulations described in  the previous Section
can be  done in the following  manner. If multiple  lensed `knots' are
detected in an image (see, e.g., Fig. 3c in Smith et al. 2005) and are
thought to be star-forming regions,  the flux will be highly dominated
by young O-stars. In principle, since young massive stars are rare due
to their evolutionary  process, only a few are  expected in these star
forming knots (an ultraviolet and  optical spectral atlas of the Small
Magellanic Cloud  includes $<$20 O-stars,  see Walborn et  al.  2000).
Observing  these areas,  e.g.   in the  UV  band, which  characterizes
regions of  star formation, with periodic  photometry, the variability
of  the  observed light  curves  will be  related  to  the number  and
separation of  these stars  present in the  star forming  regions (the
contamination by late-type  stars in the UV will  be almost null).  In
practice, one could treat  the problem statistically by simulating the
observed variability and thus put  limits on the amount and luminosity
of  young dominating  stars.  Knowing  the luminosity  of  these stars
accurately  is  important, because  their  masses  derived by  stellar
evolutionary models and by  stellar atmosphere models can be compared.
Stellar  evolution  theory  and   initial  mass  function  might  take
advantage of these results  as well.  Gravitational microlensing might
be  the  only  tool to  `resolve'  these  stars  in clusters  of  star
formation,  otherwise   impossible  to  investigate   in  galaxies  at
moderate/high redshift.

Spectroscopy  of microlensed  star-forming  regions might  help to  put
limits on  their nature as  well. Gravitational microlensing  of broad
spectral lines in  QSOs has been studied theoretically  by a number of
authors (e.g., Abajas et al. 2002, Lewis et al.  2004, Richards et al.
2004) and used  to put limits on  the size of the  broad line emitting
regions of the  QSOs.  In those cases, the natural shape  of a line is
distorted by the  complex net of caustics produced  by the microlenses
on the source  plane.  Since microlensing of the  broad line region is
expected when its physical size is of the order of the Einstein radius
of  the lens  projected onto  the source  plane,  microlensed spectral
lines give an idea of such physical sizes.  In the same way, observing
typical  spectral lines  of O-stars  (e.g.  in  the UV,  O  {\small V}
$\lambda$1371\AA, C  {\small III} $\lambda$1176\AA~in  the optical, He
{\small  II}   $\lambda$4686\AA,  N  {\small  III}$\lambda$4634\AA~and
$\lambda$4640\AA) one  would expect  the lines to  be deformed  by the
presence of the caustics (due to magnifications/demagnifications), and
these  variations in  the spectral  lines  might reveal  the size  and
populations of the star forming regions.


\begin{figure}[ht]
\epsscale{1.00}
\plotone{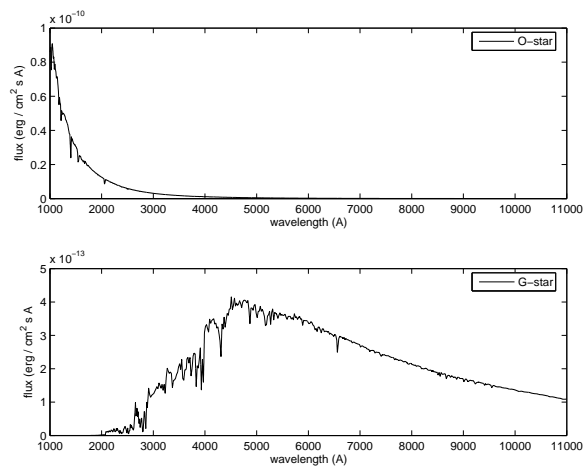}
\caption{Spectra of  O-star and G-star  models. The contribution  of a
solar-type star to  the total flux in the (far)UV  band can be ignore,
since  it  is  several  others  of  magnitudes  lower.  See  text  for
references.}
\label{fig4}
\end{figure}

\begin{figure}[h]
\epsscale{1.00}
\plotone{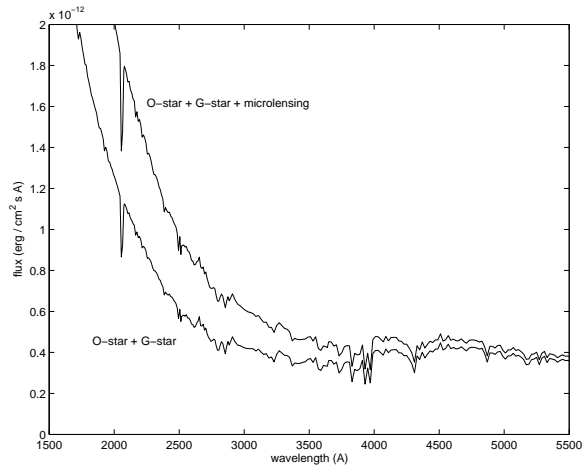}
\caption{The  toy model for  a star  formation region  is made  of two
superimposed  spectra of  a O-star  and  a G-star.  The relative  flux
between the  two types of  stars is 10\%  and 90\% for the  O-type and
G-type  respectively.  We  introduce  the microlensing  effect  as  an
enhancement  in the  flux of  the O-stars  by a  factor of  1.6, which
corresponds to a decrease of 0.5 magnitudes (upper line).}
\label{fig5}
\end{figure}
\begin{figure}[ht]
\epsscale{1.00}
\plotone{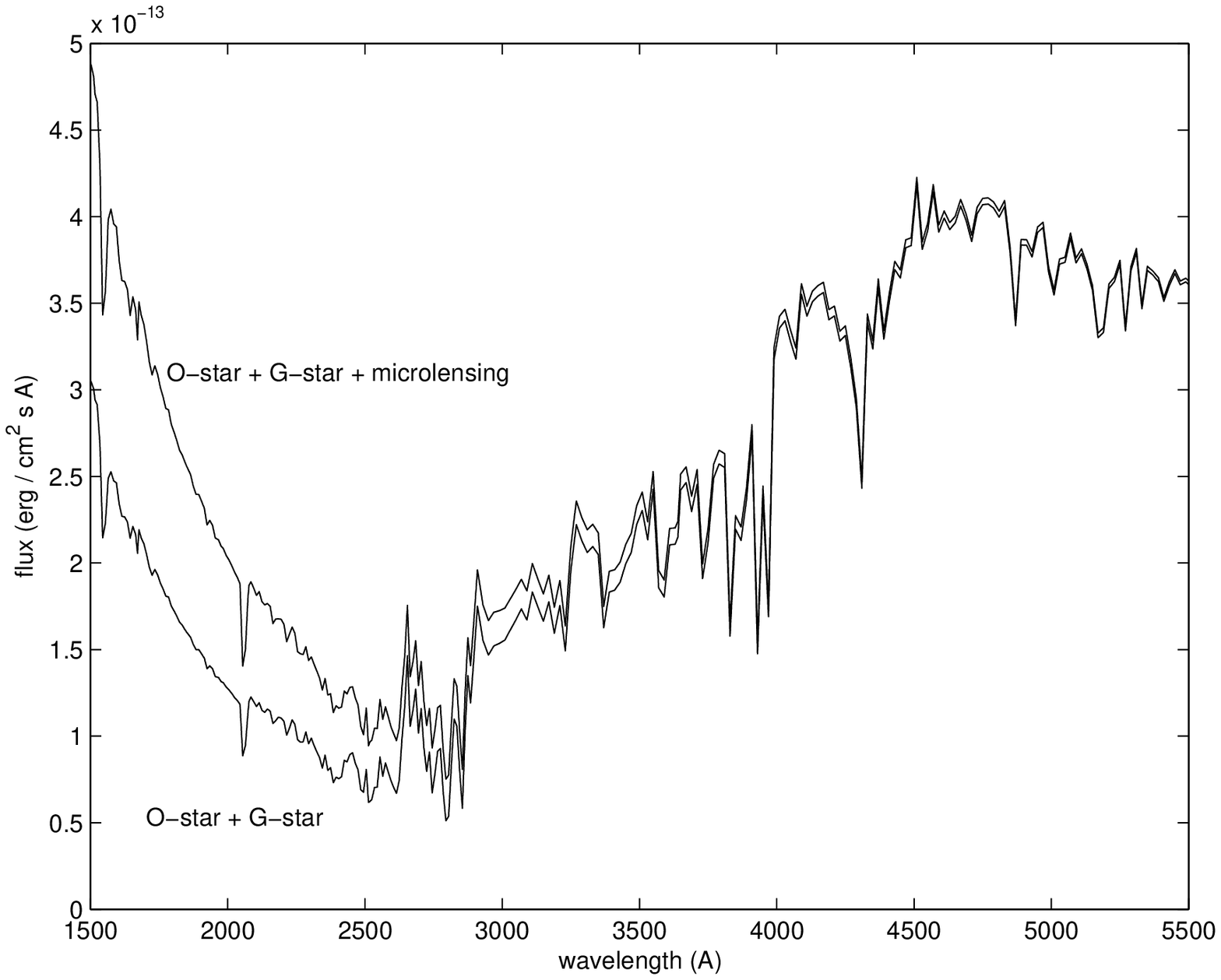}
\caption{As Fig.~\ref{fig5} but  the flux contribution is 1\% and
99\% from the O-star and the G-star respectively.}
\label{fig6}
\end{figure}
\begin{figure}[ht]
\epsscale{1.00} \plotone{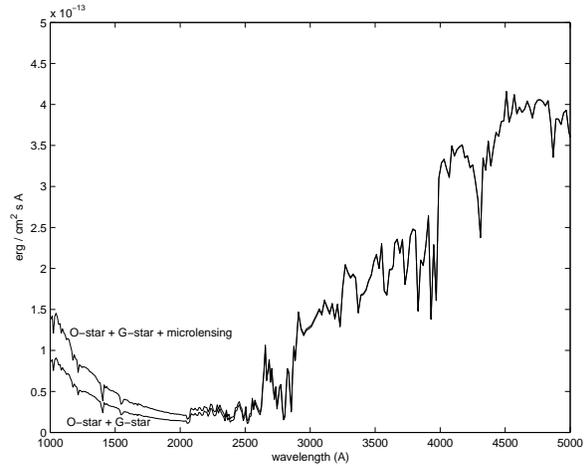}
\caption{As  Fig.~\ref{fig5} but  the flux contribution  is 0.1\%
and 99.9\% from the O-star and the G-star respectively.}
\label{fig7}
\end{figure}

To  illustrate this,  we plot  in  Fig.~\ref{fig4} the  spectra of  an
O-star (upper pannel) and a solar-like G-star (lower pannel), obtained
from                 the                 Kurucz                 models
database\footnote{http://garnet.stsci.edu/STIS/stis\_models.html}. For
any  unresolved  star formation  region,  the  (far-)UV  range of  the
spectrum will be dominated by these O-stars. Late-type stars fluxes in
UV  are  several  orders of  magnitude  lower  and  thus they  do  not
contribute  significantly  to  the   total  luminosity  and  the  flux
distribution  in  the  upper  pannel  of Fig.~\ref{fig4}  might  be  a
representative  one for  that part  of the  spectrum  (different lines
might be present, obviously).  Microlensing affecting this part of the
spectrum  will only  show an  enhancement  of the  flux. However,  the
effect  is  slightly  different   when  using  optical  range  of  the
spectrum. In this  case, the flux contribution due  to late-type stars
starts to  be dominant, although  O-stars flux is  still significantly
present. We construct  a toy star formation region  model,
merging  the  spectra  of   the  O-star  and  the  G-star  shown  in
Fig.~\ref{fig4},  assuming that  90\%  of the  total  flux comes  from
solar-like stars and the rest is produced by early-type stars. The toy
model is  depicted in Fig.~\ref{fig5} (lower line,  spectrum marked as
'O-star$+$G-star')   showing  only   the   1500\AA-5500\AA ~wavelength
interval.   When   the  star  formation  region   travels  across  the
magnification pattern in Fig.~\ref{fig1},  late-type stars will act as
a constant  flux background  as a whole  and microlensing  will affect
mainly O-stars. In this way, the microlensing signature in the spectra
will be  a flux  ratio variation between  O-stars and  late-type stars
spectral  lines. This is  shown also  in Fig.~\ref{fig5}  (upper line,
spectrum  marked as  'O-star$+$G-star$+$microlensing').  There  is not
only an  enhancement of the flux  in the bluest part  of the spectrum,
but  also  a  deformation  of  certain  lines  due  to  the  different
microlensing effect on the different type of stars. In Fig.~\ref{fig6}
and Fig.~\ref{fig7} we repeat  the procedure, but assuming a different
relative flux between the two  types of stars. In Fig.~\ref{fig6}, 1\%
of the  flux is coming  from O-type stars  and 99\% is  from late-type
stars; in Fig.~\ref{fig7} the percentage is 0.1\% and 99.9\% for early
and late-type  stars respectively.  As shown  in Fig.~\ref{fig3}, high
variability is  expected. Both Figures~\ref{fig2}  and \ref{fig3} show
that  the amount  of variability  depends on  the nature  of  the star
formation  regions (number  of stars,  size of  the  regions...). This
means that comparing several consecutive  spectra one would be able to
statisticaly determine  the relative flux variability  of the spectral
lines  and continuum due  to the  presence of  the caustics,  and thus
compare them with the expected one from the simulations, puting limits
to the number and distribution of the early-type stars.


A key  point to note is  that to detect these  microlensing effects on
star formation regions  is the time scale of  the events.  Considering
the lens  configuration describe in Sec.~\ref{sim}, we  can estimate a
typical separation between microcaustics  in the magnification maps in
Fig.~\ref{fig3}.   This  separation is  $\sim$0.005~ER  for the  upper
right-hand  panel, which  corresponds  to approximately  10$^{-3}$~pc.
Assuming   a   transverse   velocity   for  the   source   galaxy   of
$\sim$6000~km/s (see Kayser et al. 1986), the resulting time-scale for
the events is $\sim$50 days. The time-scales of the events get shorter
when the number of O-stars  gets higher, although the flux variability
is smaller. This means that six data points in a time period of around
three  months should  be able  to  described the  type of  variability
involved in the gravitational microlensing scenario.

\section{Conclusions}
We  described   in  this  contribution  how   to  apply  gravitational
microlensing  to  the observations  of  unresolved extragalactic  star
forming  regions.   The discussion  shows  that  due  to the  caustics
configuration  in   the  magnification  maps  of   the  region,  rapid
monitoring campaigns, both  photometric or spectroscopic, would reveal
high  variability  fluctuations  due   to  the  number  of  early-type
stars. The specific amount of variability will depend on the number of
stars and  their distribution in the  region, as well as  on the exact
configuration  of the microlenses  in the  lensing galaxy.   Thus, the
study of a particular system requires the knowledge of a lens model to
perform the right  simulations and the analysis of  the results should
be based on  a statistical approach.  The advantage  of the method, if
these  circumstances  take  effect,  is  that  we  might  be  able  to
investigate star formation regions which are difficult to analyse with
more traditional techniques.

\acknowledgments

 \end{document}